\documentclass[preprint,12pt]{elsarticle}

\usepackage{amssymb}
\usepackage{amsmath}

\usepackage{subcaption}
\usepackage{hyperref}
\usepackage{cleveref}
\usepackage{lineno}
\usepackage{upgreek}

\usepackage{siunitx}
 \DeclareSIUnit\bar{bar}

\DeclareSIUnit\liter{\ell}

\usepackage[acronym,xindy,toc,nonumberlist]{glossaries}

\newacronym{tlk}{TLK}{Tritium Laboratory Karlsruhe}
\newacronym{tapir2}{T$_2$ApIR}{Tritium Absorption InfraRed Spectroscopy 2}

\newacronym{lara}{LARA}{Laser Raman}
\newacronym{muRa}{$\upmu$Ra}{micro Raman}

\newacronym{ftir}{FTIR}{Fourier-transform infrared}

\makenoidxglossaries

\newcommand{\requiredTritiumAmountGram}{\SI{14}{\gram}}

\newcommand{\requiredTritiumAmountToFillCellGram}{\SI{\approx 1.5}{\gram}}
\newcommand{\requiredTritiumAmountToFillCellDecayHeatWatt}{\SI{\approx 0.5}{\watt}}

\newcommand{\requiredTemperatureLowerLimitKelvin}{\SI{10}{\kelvin}}
\newcommand{\achievedTemperatureLowerLimitKelvin}{\SI{12}{\kelvin}}

\newcommand{\operatingPressureLimit}{\SI{2.5}{\bar\,a}}
\newcommand{\designPressureLimit}{\SI{5}{\bar\,a}}

\newcommand{\achievedTemperatureSecondStageLowerLimitKelvin}{\SI{6}{\kelvin}}   
\newcommand{\achievedTemperatureSecondStageUpperLimitKelvin}{\SI{80}{\kelvin}}  

\newcommand{\tritiumBoilingPoint}{\SI{24.99}{\kelvin}} 
\newcommand{\tritiumTriplePoint}{\SI{20.62}{\kelvin}} 

\newcommand{\hydrogenEquilibratedTriplePoint}{\SI{13.80}{\kelvin}}  
\newcommand{\hydrogenTriplePoint}{\hydrogenEquilibratedTriplePoint{}}    

\newcommand{\betadecay}{$\upbeta$-decay}
\newcommand{\orthopara}{ortho/para}

\newcommand{\orthoparaconverter}{ortho/para-converter}
\def\HTwo{H$_2$} 
\def\DTwo{D$_2$} 
\def\TTwo{T$_2$} 
\def\QTwo{Q$_2$}

\journal{Fusion Engineering and Design}

\begin{document}

\begin{frontmatter}



\title{Commissioning of an experiment for thermodynamic and spectroscopic studies of hydrogen isotopologues at cryogenic conditions }

\author[inst1]{Joshua Kohpeiß}
\author[inst1]{Dominic Batzler}
\author[inst1]{Beate Bornschein}
\author[inst1]{Lutz Bornschein}
\author[inst1]{Robin Größle}
\author[inst1]{Daniel Kurz}
\author[inst1]{Ralph Lietzow}
\author[inst1]{Alexander Marsteller\texorpdfstring{\corref{cor1}}} \ead{alexander.marsteller@kit.edu}
\author[inst1]{Michael Sturm}
\author[inst1]{Stefan Welte}

\affiliation[inst1]{organization={Institute for Astroparticle Physics, Tritium Laboratory Karlsruhe (IAP-TLK), Karlsruhe Institute of Technology (KIT)},
            addressline={Hermann-von-Helmholtz-Platz 1}, 
            city={Eggenstein-Leopoldshafen},
            postcode={76344}, 
            state={Baden-Württemberg},
            country={Germany}}
            
\cortext[cor1]{corresponding author}

\begin{abstract}
To study thermodynamic properties and dynamic phase space behavior of hydrogen isotopologues (\QTwo{}) at cryogenic temperatures and at high density, the \gls{tapir2} experiment has been set up and commissioned at \gls{tlk}. 
In the frame of the experiment, \QTwo{} behavior in different phases, \orthopara{} states, temperatures (\requiredTemperatureLowerLimitKelvin{} - \SI{300}{\kelvin}) and pressures (up to \operatingPressureLimit{}) will be investigated with optical methods, infrared and Raman spectroscopy. 
The facility consists of a fully tritium compatible cryostat, which includes an optical cell, \orthopara{} converter and windows for optical and spectroscopic studies. 
The cryostat can be cooled below the \HTwo{} triple point by a two-stage cryocooler and contains openings in the cryogenic shielding for the optical access. 
The challenge of combining these scientific requirements in a design with high amounts of tritium (\requiredTritiumAmountGram{}), in a limited space, all while maintaining the \gls{tlk} safety philosophy was solved by the presented design. 
The experiment is ready to be fully integrated into the \gls{tlk} closed loop tritium infrastructure. This contribution reports a comprehensive overview of the commissioning phase of the experimental facility and the results of the first commissioning experiments, including cryogenic performance tests, commissioning experiments with non-radioactive gases, and tests of the analytical instruments. 

\end{abstract}


\begin{highlights}
\item Tritium compatible cryostat for optical investigations
\item FTIR and Raman spectroscopy of cryogenic hydrogen isotopologues
\item Crystal growth of cryogenic hydrogen isotopologues
\end{highlights}


\end{frontmatter}


\glsresetall{}

\section{Introduction}
\label{sec:introduction}

Solid and gaseous tritium are used for direct neutrino mass determination \cite{Bornschein2008,Aker2025,Esfahani2017,Amad2025} while the deuterium-tritium fusion reaction is the most promising contender for fuel in nuclear fusion for electric power  \cite{VanOost2023}. 
Current research at \gls{tlk} focuses on the thermodynamic properties of high purity H-D-T-mixtures \cite{Niemes2023,Wydra2023,Priester2023,Wydra2025} and the development of analytical tools \cite{Priester2022,Niemes2021,Aker2020,Sturm2009,Priester2017} for target production, pellet production, cryogenic pumping, cryogenic distillation, and water detritiation \cite{Cristescu2017} for inertial and magnetically confined fusion. 

In cryogenic distillation for isotope separation, the six hydrogen isotopologues (\QTwo{} = \TTwo{}, DT, \DTwo{}, HT, HD, and \HTwo{}) are separated in a cryogenic refraction column via the differences in their vapor pressures at a given temperature. 
As the highest boiling isotopologue, with a boiling point of \tritiumBoilingPoint{} at \SI{1}{bar} \cite{Souers1986}, tritium accumulates at the bottom of the column in the liquid phase.
To monitor the concentration of \TTwo{}, as well as residual amounts of HT and DT, a measurement system is required. 
At the \gls{tlk}, infrared (IR) absorption spectroscopy is under investigation as an online and inline monitoring tool and has been successfully calibrated for the inactive isotopologues with an accuracy of better than \SI{5}{\percent} absolute \cite{Groessle2017}.  

The production of fuel targets for inertial confined fusion for a commercial fusion power plant requires a fast rate of target production in order to achieve the desired output power while keeping overall tritium inventory low \cite{Goodin2006}.
A key step in this process is loading targets with a deuterium-tritium mixture. 
Depending on the target concept the latter is solidified by freezing it out .

In order to address these open questions an experimental setup to calibrate spectroscopic methods, study thermodynamic properties, and investigate the dynamic phase space behavior of all six hydrogen isotopologues (\QTwo{}) at cryogenic temperatures at high density is necessary. 
Such a setup should allow investigating all six hydrogen isotopologues (\QTwo{}) in different phases, mixtures, pressures, \orthopara{} states and at the triple point using infrared and Raman spectroscopy as well as optical analysis. 
Based on previous experience with the inactive hydrogen isotopologues (\HTwo{}, HD, \DTwo{}) \cite{Groessle2017,Groessle2015} the \gls{tapir2} experiment \cite{Krasch2020} has been set up and commissioned at the \gls{tlk}. 

In this work, the as-built experimental setup is described and the performance of key parameters that could be achieved during commissioning of the experiment are presented. 
\section{Experimental idea}
\label{sec:experimental_idea}

\begin{figure}[t]
\centering
\includegraphics[width=\textwidth]{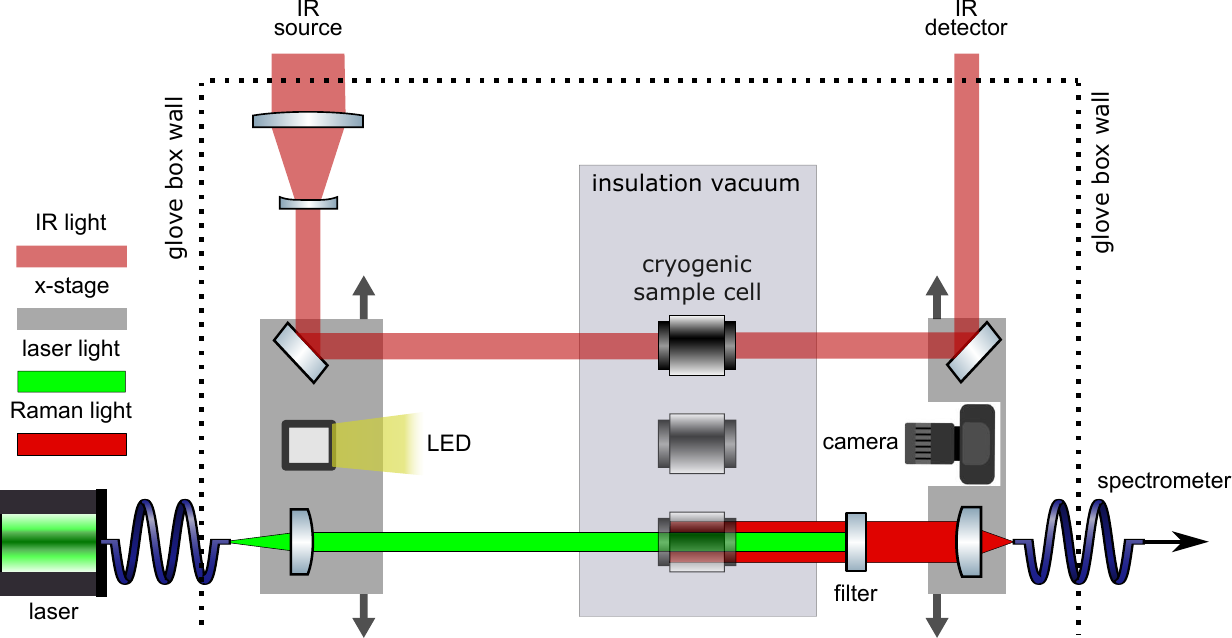}
\caption{
Experimental idea including IR-system, throughpass laser Raman-system and photography in a measurement cell.
The sample cell is included in the figure multiple times to illustrate the photography and Raman beam path.
Only one cell at a fixed position is present in the setup.
}
\label{fig:optical_sketch}
\end{figure}

The principal goal of the \gls{tapir2} experiment is the spectroscopic and optic investigation of all six hydrogen isotopologues and three spin isomers (\orthopara{}) in the liquid and solid phase, covering the temperature range from room temperature down to cryogenic temperatures.
In particular, the spectroscopic techniques of interest are transmission infrared absorption spectroscopy and laser Raman spectroscopy, which have both been successfully used with the inactive hydrogen isotopologues (\HTwo{}, HD, \DTwo{}) \cite{Groessle2017, Groessle2015} or gaseous tritiated isotopologues (HT, DT, \TTwo{}) \cite{Aker2020, Sturm2009}.
The conceptual setup combining a cryogenic sample cell with a transmission IR beam path and throughpass Raman arrangement is depicted in \cref{fig:optical_sketch}. 
Where Raman spectroscopy is meant to be used to mainly probe the molecular state and composition, IR includes additional information about the molecular interactions like collisions, phonons and molecular dimers \cite{Groessle2020}.

Complementary to and aided by the spectroscopic and optic investigations of the cell content, the experiment aims at exploring the dynamic behavior during phase change.
For this purpose, the experimental goal is to cover as large a fraction of the phase space as is reasonably achievable within the technological restraints.

The biggest technical restraints on top of the technical requirements necessary for achieving the scientific goals, is that the setup needs to be compatible with tritium operation \cite{Welte2015} and the Pressure Equipment Directive\footnote{EU Pressure Equipment Directive \url{http://data.europa.eu/eli/dir/2014/68/oj} (accessed 21. November 2025)}.
Combining the necessary conditions for safe operation with the above scientific requirements posed several design challenges, to which the solution is presented in the following.

The \gls{tapir2} experiment, a fully tritium compatible test facility including a cryostat with interfaces for IR- and Raman spectroscopy as well as optical access, was designed, set up and tested. 
The cryostat is connected to the primary system including a gas supply, process loop, vacuum system and tritium infrastructure for tritium transfer and tritium waste gas handling \cite{Krasch2020,Mirz2017}. 
To fulfill the TLK safety philosophy, the facility has to be integrated into a glove box with auxiliary systems such as the underpressure- and the tritium retention system \cite{Welte2015}.

\section{Tritium-compatible cryostat for optical investigation }
\label{sec:cryostat_setup}

In designing the experimental setup \cite{Krasch2020}, several partially conflicting technical requirements needed to be taken into account.
In order to achieve the scientific goals, the cryogenic sample cell needs to be able to reach temperatures below the triple point of protium (\HTwo{}) at \hydrogenTriplePoint{} in order to be able to solidify all six hydrogen isotopologues (\QTwo{}).
The temperature also needs to be regulated and stabilized so the liquid and solid phase space of the different isotopologues can be covered.
Additionally, to cover a wider range of the phase space, the setup was designed to handle overpressures of up to \operatingPressureLimit{}.

\begin{figure}[t]
\centering
\includegraphics[width=\textwidth]{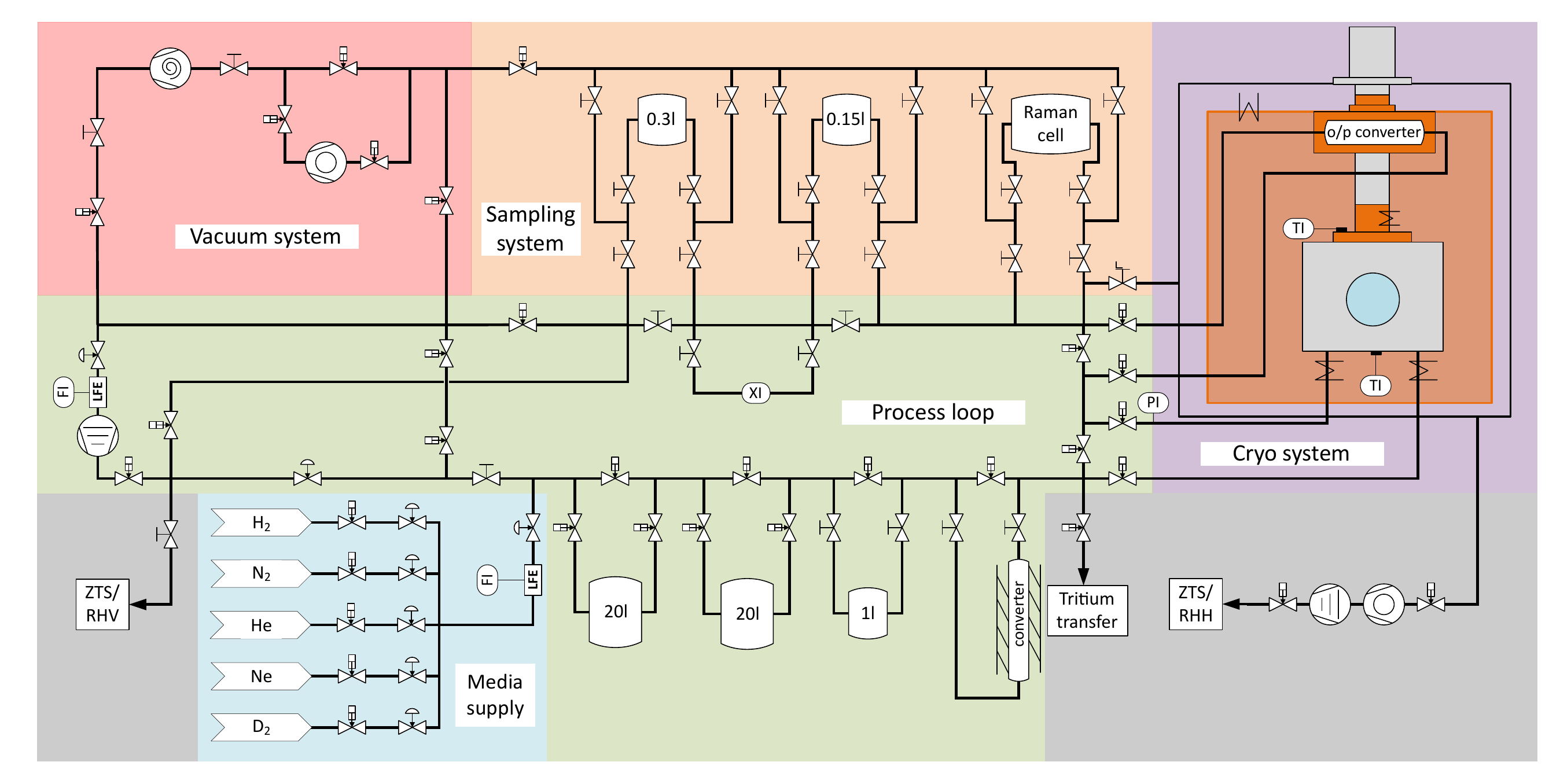}
\caption{Simplified flow diagram of the \gls{tapir2} primary system (optical system depicted in \cref{fig:optical_sketch} not shown).}
\label{fig:simplified_flow_diagram}
\end{figure}

As the spin isomers of the homonuclear isotopologues (\HTwo{}, \DTwo{}, \TTwo{}) lead to different spectroscopic properties, and the natural conversion between the ortho and para states is very slow, the setup should be able to convert these states to the equilibrium at a given temperature.
For this purpose, a cryogenic \orthoparaconverter{} \cite{Krasch2023} is included in the setup.
In addition to the converter, the setup contains a room temperature chemical catalyst for spin isomer conversion as well as production of heteronuclear isotopologues. 
A \gls{lara} system at room temperature serves as a reference analytical system to measure the isotopologue and isomer concentration.
All of these components are arranged in a circulation loop with buffer vessels to ensure that homogeneous gas mixtures can be produced.
The simplified flow diagram of this setup is shown in \cref{fig:simplified_flow_diagram}.

The requirement for stable, low temperature stands in contrast with the desire to have good optical access via two windows for transmission spectroscopy, which require a sight line to room temperature, increasing the heat load on the cryogenic cell.
To decrease other heat loads on the cell, a radiation shield is used to prevent the majority of room temperature radiative heat from reaching the cell.

On top of these technical requirements, the setup needs to be compatible with tritium operation.
In general, this means adherence to the \gls{tlk} administrative and technical framework for isotope laboratory operation \cite{Welte2015}.
Among these regulations is a fully metal, glass or ceramic sealed primary system containing the tritium.
This primary system is integrated into the closed tritium loop of the \gls{tlk} including a tritium transfer system, and tritium waste gas handling.
The entire setup is surrounded by a secondary containment enclosure, the atmosphere of which is continuously decontaminated.
The secondary containment enclosure imposes limitations on the size of the cryostat which needed to be considered during the design process.

All of those requirements were met in the previous design \cite{Krasch2020}. 
The cryostat consists of a measurement cell with optical windows, an \orthoparaconverter{}, temperature and pressure sensors, heaters, a copper connection to a two stage cryocooler, and a thermal radiation shield. 
All components are contained in a DN\,160\,CF cube extended with a pipe and a DN\,160\,CF cross with a design pressure of \designPressureLimit{} and a volume of \SI{16}{\liter}.
The measurement cell and the converter combined displace a volume of \SI{\approx 20}{\milli\liter} inside the insulation vacuum. 

\begin{figure}[t]
\centering
\includegraphics[width=0.9\textwidth]{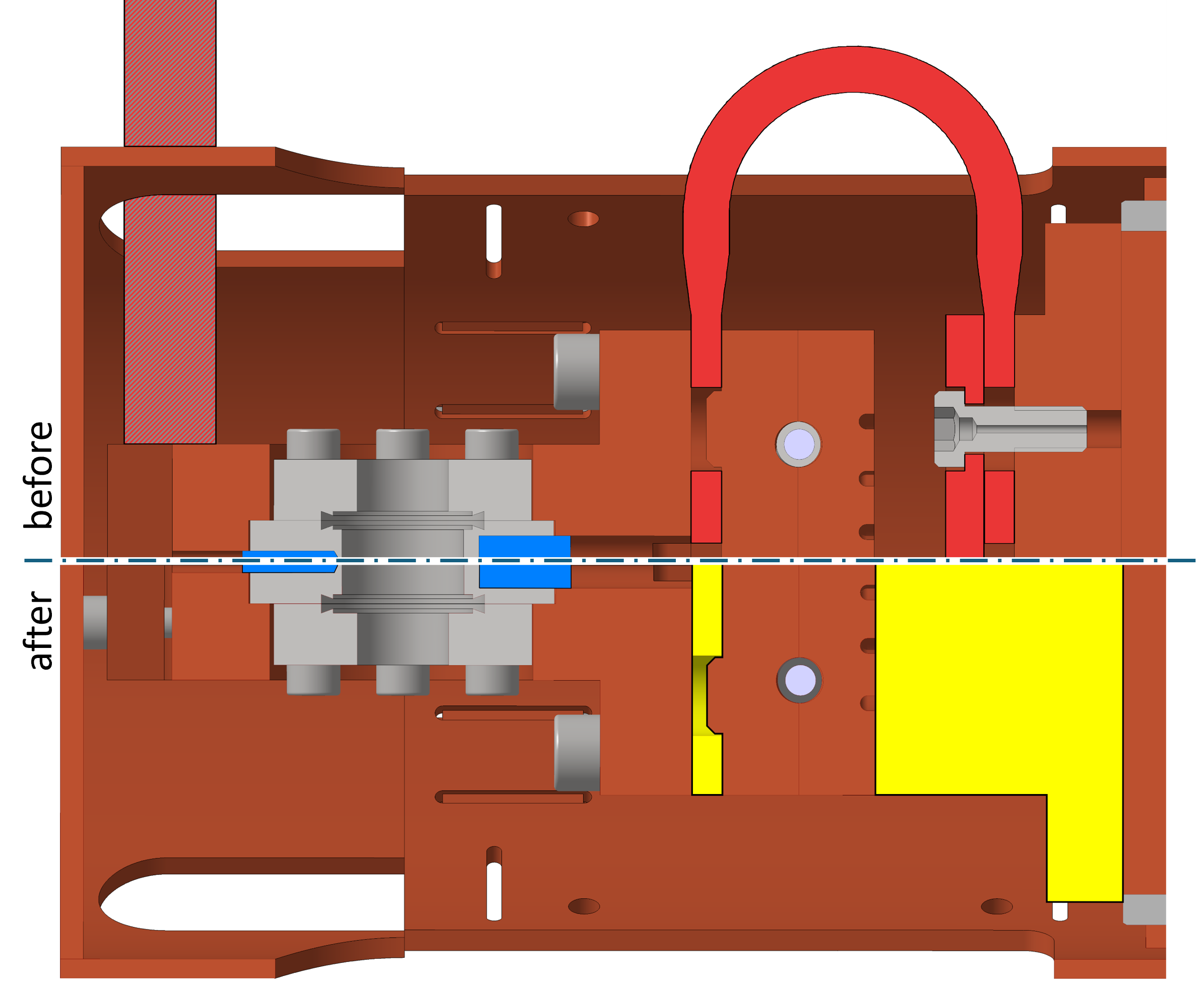}
\caption{
Cryostat design changes. 
The top half shows the old design, the bottom half the new design.
ZrO$_2$ rods and copper ribbons were removed (red) and replaced by a solid copper block (yellow). 
Heat is conducted to a cold head (not shown) on the right. 
TVO sensor and Pt1000 of the cryogenic measurement cell (grey) are shown in blue.
}
\label{fig:design_changes}
\end{figure}

During commissioning, the set-up was iterated over in several disassembly and assembly steps, increasing the thermal connection between the second stage of the cryocooler and the cryostat insert components (see \cref{fig:design_changes}):
\begin{itemize}
    \item The copper ribbons were replaced by a solid copper block, which improves the thermal connection between the insert and the second stage of the cryocooler as well as the positioning and stability of the optical cell.
    However, this comes at the the cost of flexibility and distance to the (heated) converter.
    \item The ZrO$_2$ rods were no longer required and thus removed.
    \item Apiezon\textsuperscript{\tiny\textregistered}  N was used to improve the thermal connection between several components and to the cryocooler. 
    \item The heat shield was extended with aluminum tape to further improve insulation against heat radiation. 
    \item To improve the thermal connection between the copper parts, threads were strengthened using wire thread inserts to enable a higher torque of screws.
\end{itemize}
After these improvement steps, it was possible to reach a minimum process temperature of \achievedTemperatureLowerLimitKelvin{}. 
This is sufficient to cover the necessary temperature range to generate solid protium (below the triple point of protium at \hydrogenTriplePoint{}).
    
The results of the commissioning phase are presented in the following.

\section{Commissioning: cryo-performance and safety}
\label{sec:commissioning}

The commissioning of the facility can be divided into several objectives: The commissioning of the primary system to ensure safe tritium operation, cryostat commissioning to verify that the required temperatures to condense and solidify all hydrogen isotopologues can be reached, commissioning of the room temperature chemical catalyst and cryogenic \orthoparaconverter{}, and the commissioning of the optical and spectroscopic measurement methods.

\subsection{Primary system commissioning}
\label{subsec:primary_system_commissioning}

For the primary system, all functionality, operability and safety tests of mechanical and electrical components as well as their communication had to be defined and checked, including the following topics \cite{Welte2015}.
A thorough documentation in accordance with the TLK safety philosophy, including a HAZOP study, safety documentation, design approval, operator instructions, and quality assurance documentation was created and maintained throughout the design, construction, and commissioning phase. 
A significant part of this documentation contains the functionality checks of components in the process field, including the leak tightness, combined functionality, as well as normal and fail safe operation of valves, heaters, sensors, and pumps.
Safety checks regarding electrical safety, process safety and all required pressure tests (according to PED) were performed and documented.

After iterating over the design and successfully demonstrating the proper function of all components of the primary system, cryogenic commissioning was performed using the inactive hydrogen isotopologues.

\subsection{Cryostat commissioning}
\label{subsec:cryostat_commissioning}

\begin{figure}[t]
\centering
\includegraphics[width=\textwidth]{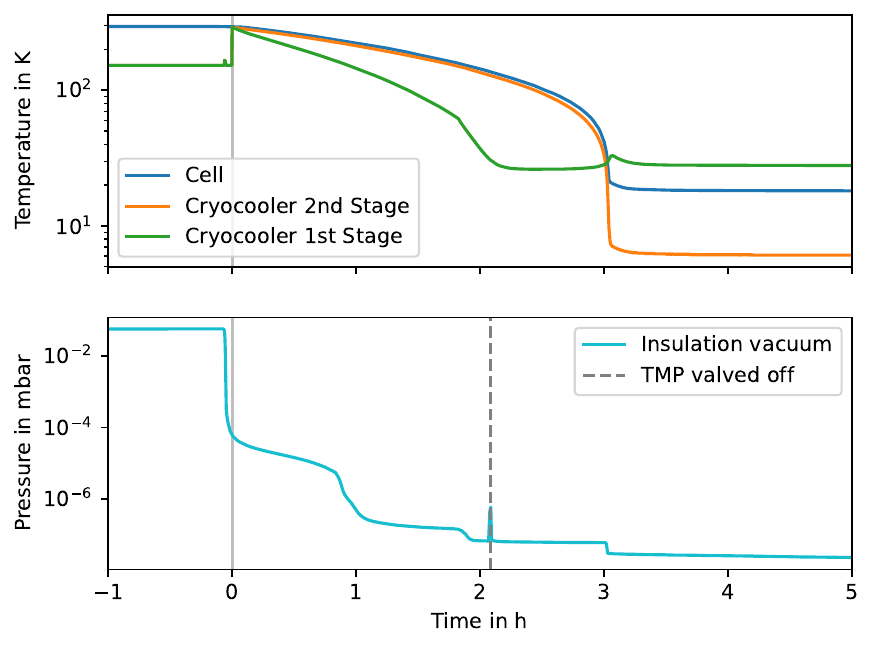}
\caption{Cooldown curve of the cryostat.}
\label{fig:cryostat_cooldown}
\end{figure}

The cryostat commissioning was performed by cooling down the cryostat with an evacuated measurement cell. 
A typical cooldown curve is shown in \cref{fig:cryostat_cooldown}. 
After about \SI{3}{\hour} the cell reaches its lowest steady state temperature.
At the start of the cooldown, the insulation vacuum is pumped using a turbo molecular pump. 
After around \SI{2}{\hour}, the turbo molecular pump can be valved off as the cold surfaces begin to function as a cryogenic pump. 
This allows for the pressure in the insulation vacuum of the cryostat to reach on the order of \SI{1E-9}{\milli\bar}. 

\begin{figure}[t]
\centering
\includegraphics[width=0.8\textwidth]{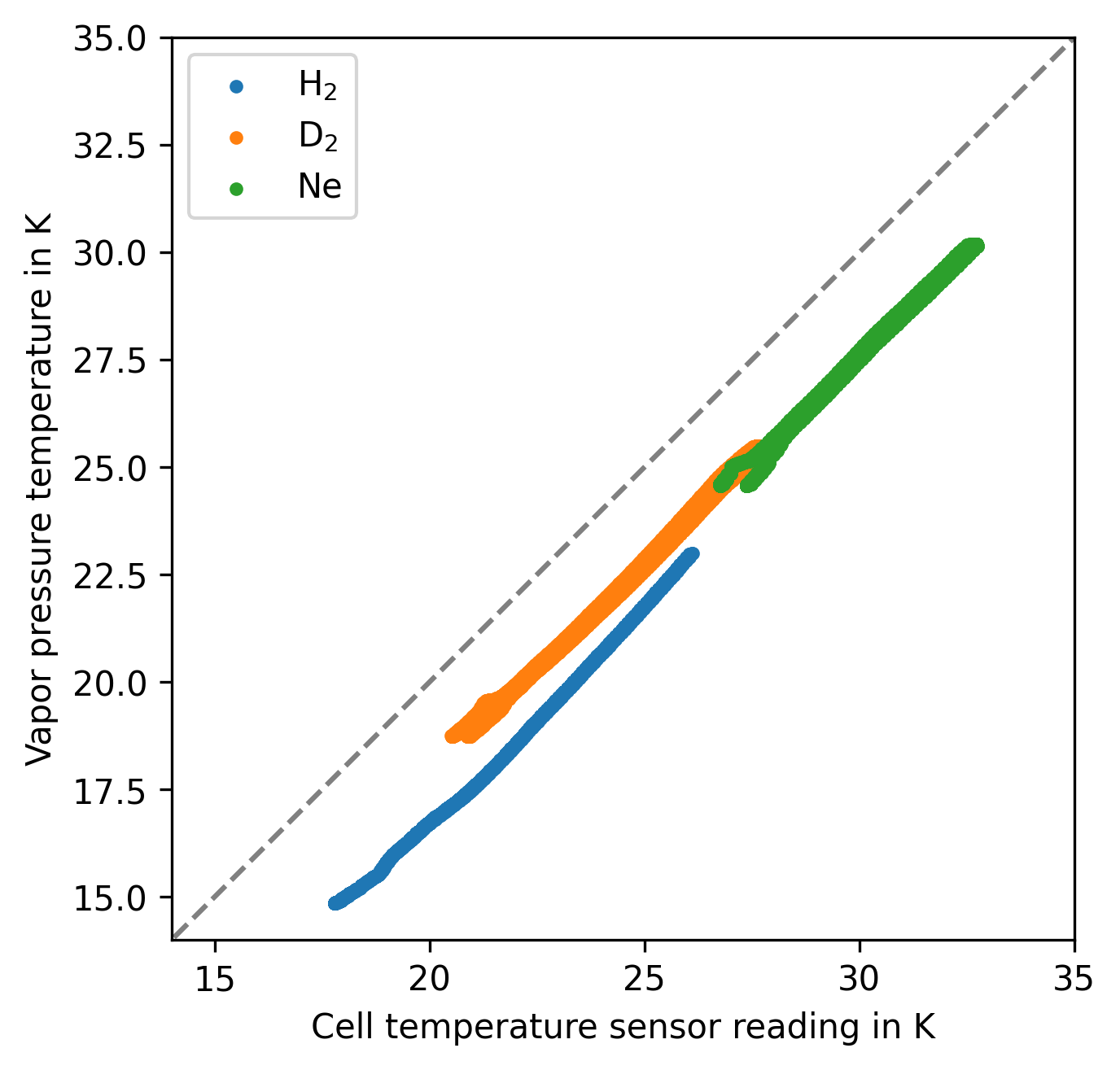}
\caption{Calibration of the cell temperature sensor using the vapor pressure derived temperature of the measurement cell contents.
The dashed line represents the ideal exact correspondence between cell temperature and sensor reading and is shown to guide the eye.}
\label{fig:temperature_calibration}
\end{figure}

After cooldown, the measurement cell was filled in several test runs with protium, deuterium, and neon. 
Each gas could be condensed into a liquid and frozen into a solid. 
Based on these test runs, a calibration of the temperature sensors attached to the measurement cell was performed (\cref{fig:temperature_calibration}). 
This was done by relating the temperature sensor reading to the temperature of the cell derived by comparing vapor pressure measurements over the liquid phase in the cell to literature saturation vapor pressure curves \cite{Souers1986, Linstrom1997}.  

\begin{figure}[t]
\centering
\includegraphics[width=\textwidth]{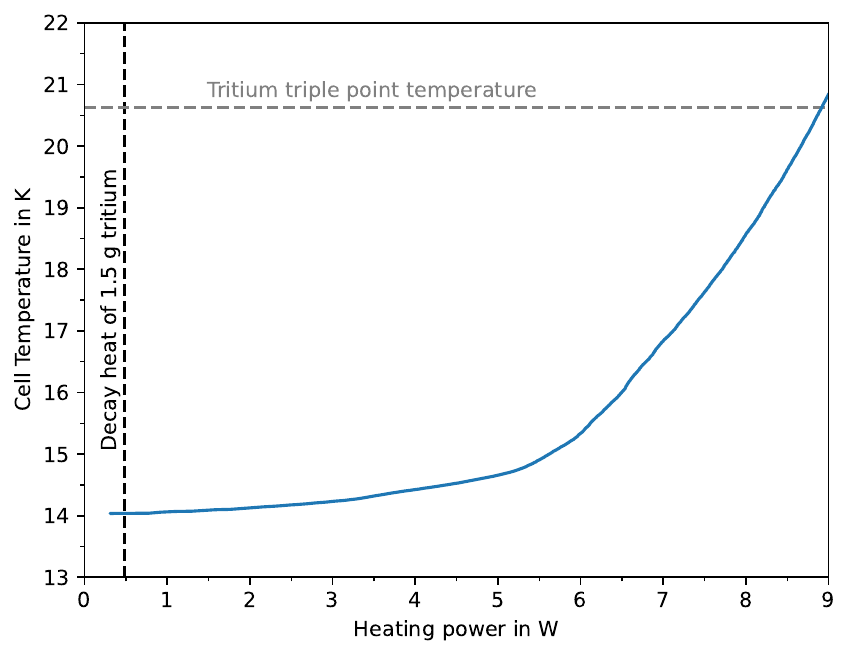}
\caption{\HTwo{} vapor pressure temperature of the measurement cell as a function of the 2nd stage heating power. The triple point temperature of tritium can easily be reached, even with a decay heat of \requiredTritiumAmountToFillCellDecayHeatWatt{} for a cell completely filled with \TTwo{}.}
\label{fig:cell_temperature_over_heater_power}
\end{figure}

Heaters on the cold stages of the cryocooler allow for temperature control of the cryogenic insert, consisting of the measurement cell and the cryogenic converter. 
Using these heaters, the temperature at the 2nd stage of the cryocooler can be varied from \achievedTemperatureSecondStageLowerLimitKelvin{} up to above \achievedTemperatureSecondStageUpperLimitKelvin{}. 
Achieving the high temperatures allows for desorption of nitrogen and oxygen, traces of which can be present as impurities, and can condense onto the windows, decreasing the performance of optical and spectroscopic measurements. 
In addition, these heaters allow the simulation of a heat load not present with the inactive hydrogen isotopologues: the decay heat of tritium. 
By supplying a defined heat load via the heaters, the expected impact on the achievable temperatures due to the decay heat load was simulated (\cref{fig:cell_temperature_over_heater_power}). 
A total amount of \requiredTritiumAmountToFillCellGram{} of tritium is necessary in order to fill the measurement cell with liquid tritium, which corresponds to a heat load of \requiredTritiumAmountToFillCellDecayHeatWatt{}. 
It could be demonstrated, that this heat load does not prevent the measurement cell from reaching the triple point temperature of tritium at \tritiumTriplePoint{}, allowing for solidification of tritium.

\subsection{Opto-spectroscopic commissioning}
\label{subsec:opto_spectroscopic_commissioning}

The \gls{tapir2} experiment contains two separate optical systems: the cryogenic measurement cell with its optical access via sapphire windows to room temperature, and a room temperature \gls{lara} measurement system \cite{Aker2020,Sturm2009}. 

The \gls{lara} measurement system is a development of \gls{tlk} and allows for the simultaneous detection of all six hydrogen isotopologues and their \orthopara{} ratio in the gas phase. 
The particular system in use at \gls{tapir2} was previously used in an experimental setup to investigate the performance of hydrogen \orthopara{} catalysts \cite{Krasch2023}. 
Therefore, no renewed commissioning of this system was necessary. 

\begin{figure}
\centering
\includegraphics[width=\textwidth]{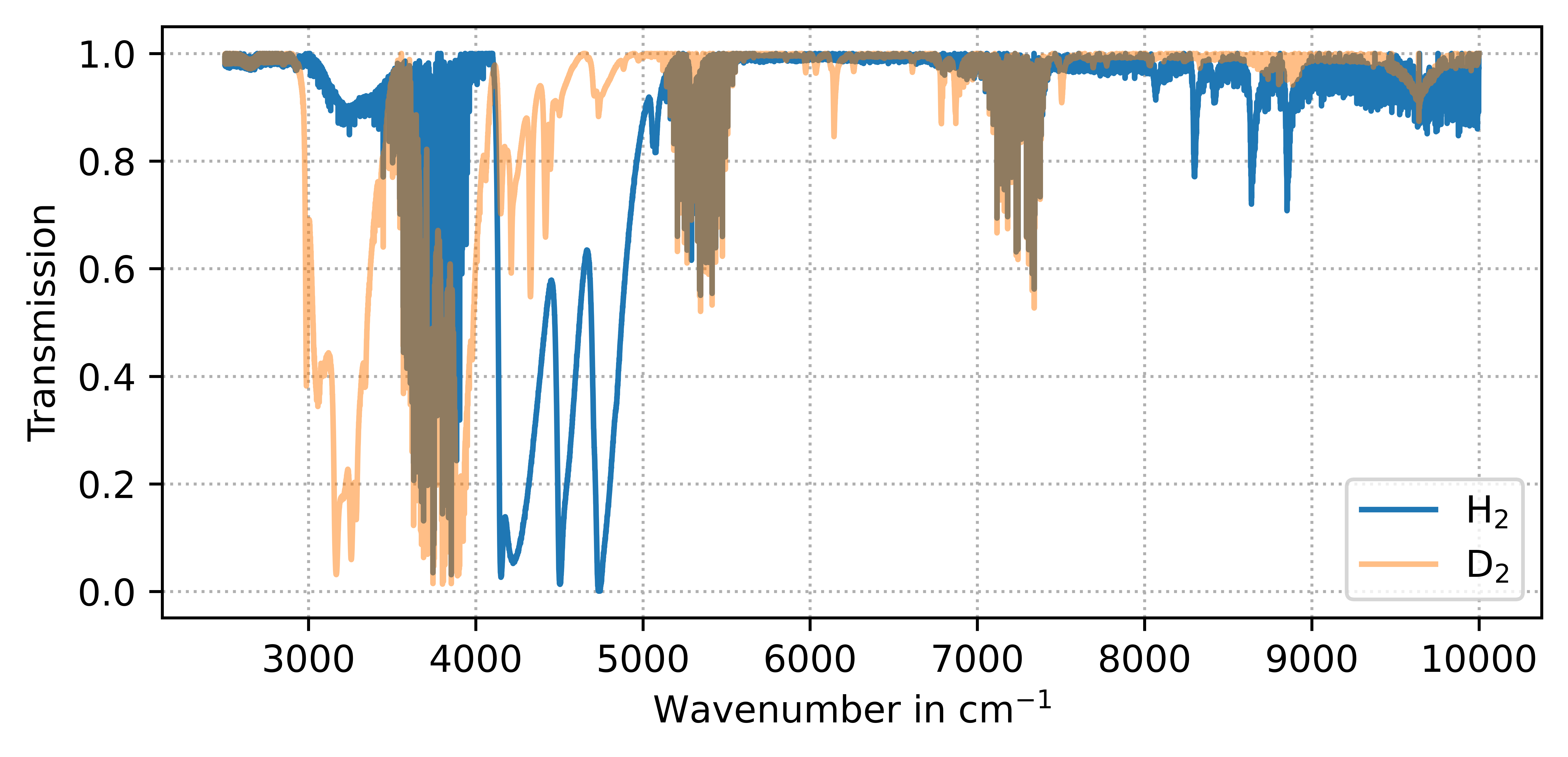}
\caption{Sample transmission infrared spectra of pure liquid \HTwo{} and pure liquid \DTwo{}.}
\label{fig:ir_spectrum}
\end{figure}

Optical access to the cryogenic measurement cell is facilitated by two optical benches outside of the cryostat, but inside of the second containment. 
For IR spectroscopy, a Bruker Vertex 70 \gls{ftir} spectrometer is used.
Mirrors (see \cref{fig:optical_sketch}) are placed onto the optical benches and the beam path from \gls{ftir} spectrometer to detector is adjusted to achieve maximum light intensity on the detector. 
Following that, spectra of liquid \HTwo{} and \DTwo{} were recorded with a sample spectrum each for both shown in \cref{fig:ir_spectrum}. 
In addition, several spectra of \HTwo{} were recorded over a period of time comparable to the natural \orthopara{}-equilibration time.
These measurements showed that the influence of \orthopara{} ratio on the spectral shape can be observed in the recorded infrared spectra of liquid protium. 

\begin{figure}[t]
\centering
\includegraphics[width=\textwidth]{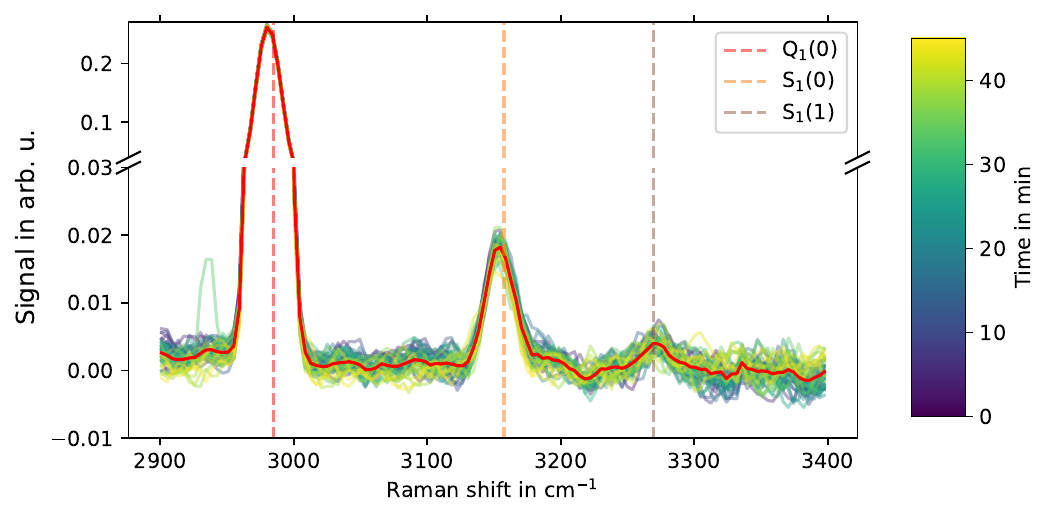}
\caption{Sample throughpass Raman spectrum of the Q$_1$-branch of liquid \DTwo{} at \SI{200}{\milli\watt} laser power and \SI{60}{\second} acquisition time per spectrum.
Literature line positions for Q$_1$(0), S$_1$(0), and S$_1$(1) lines are indicated with dashed lines.}
\label{fig:raman_spectrum}
\end{figure}

For the throughpass Raman spectroscopy system, collimating fiber couplers are placed on either side of the measurement cell.
Connected to these fiber couplers is the spectrometer and laser combination used in \gls{tlk}-developed \gls{muRa} systems in use across the \gls{tlk} \cite{Priester2022}. 
A laser cleanup filter on the excitation side removes signal from the excitation side fiber and sharpens the laser spectrum to a very narrow band around \SI{532}{\nano\meter}. 
On the collection side, one longpass filter removes the excitation laser light below \SI{534}{\nano\meter}, while a second longpass at \SI{690}{\nano\meter}filter removes nuisance photoluminescence of the sapphire windows.
The throughpass arrangement is a deviation from the standard backward arrangement in the \gls{muRa} systems across the \gls{tlk} and will require a thorough characterization.
An advantage compared to the gas phase Raman systems is the much higher density in the liquid and solid phases, allowing for much shorter measurement times, which makes it possible to investigate dynamic processes such as melting.
An example spectrum obtained with this setup is shown in \cref{fig:raman_spectrum}.

The combination of the room temperature \gls{lara} system serving as a well-understood reference measurement system with the cryogenic \gls{ftir} and throughpass Raman measurements, demonstrates that the \gls{tapir2} setup is capable of fulfilling its target of being able to perform an in-situ \orthopara{} calibration of infrared spectra as well as validating the usage of Raman spectroscopy as a monitoring tool for cryogenic tritium applications. 

For optical investigations, a \textit{Canon EOS R} system camera with a \textit{100 mm F/2.8 L IS USM macro} lens was used to take images and videos of the cryogenic measurement cell. 
It was possible to observe both condensation and solidification of \HTwo{}, \DTwo{}, and neon.
Furthermore, linear polarizers were added in front and behind the cryogenic measurement cell, showing that polariscopic investigations on solid hydrogen are possible, elucidating structural defects and stress in the freezing process. 
This is of further interest, as it allows to selectively perform spectroscopy of the solid phase in a two phase mix of solid and liquid. 
Some of the images recorded during this commissioning are shown below in \cref{sec:first_physics_results}. 

\subsection{Converter and catalyst commissioning}

The converter and catalyst were tested by \HTwo{} circulation with in-situ \orthopara{} measurement using the \gls{lara} system.
For the room temperature catalyst, a sample of protium converted to a low temperature \orthopara{} ratio via natural conversion in the liquid phase was evaporated and circulated through the loop of \gls{tapir2}, including the catalyst. 
Quick conversion of the \orthopara{} ratio to room temperature equilibrium on a time-scale of minutes was observed, demonstrating sufficient performance for the experimental needs. 

The cryogenic converter was tested in two configurations.
The first configuration was similar to that used to test the catalyst, testing conversion at room temperature using low temperature \orthopara{} ratio \HTwo{}.
In the second configuration, \HTwo{} with a room temperature \orthopara{} ratio was circulated through the catalyst at \SI{\approx20}{\kelvin}, again observing the change in \orthopara{} ratio using the \gls{lara} system. 
The observed performance of the cryogenic converter fell far below expectations in both configurations, indicating an issue with the converter material, which is subject to ongoing investigation.

\section{First physics results}
\label{sec:first_physics_results}

As part of the commissioning of the experiment, several tests and some first scientific measurements were performed: 
\begin{itemize}
    \item Various cooldowns and warmups of the system,
    \item condensing of \HTwo{}, \DTwo{}, Ne into the liquid phase,
    \item temperature ramping of \HTwo{}, \DTwo{}, Ne between the gaseous, liquid, and solid phase, and
    \item crystallization experiments with \HTwo{} and \DTwo{}.
\end{itemize}

\begin{figure}[t]
\centering
\begin{subfigure}{.3\textwidth}
  \centering
    \includegraphics[width=\linewidth]{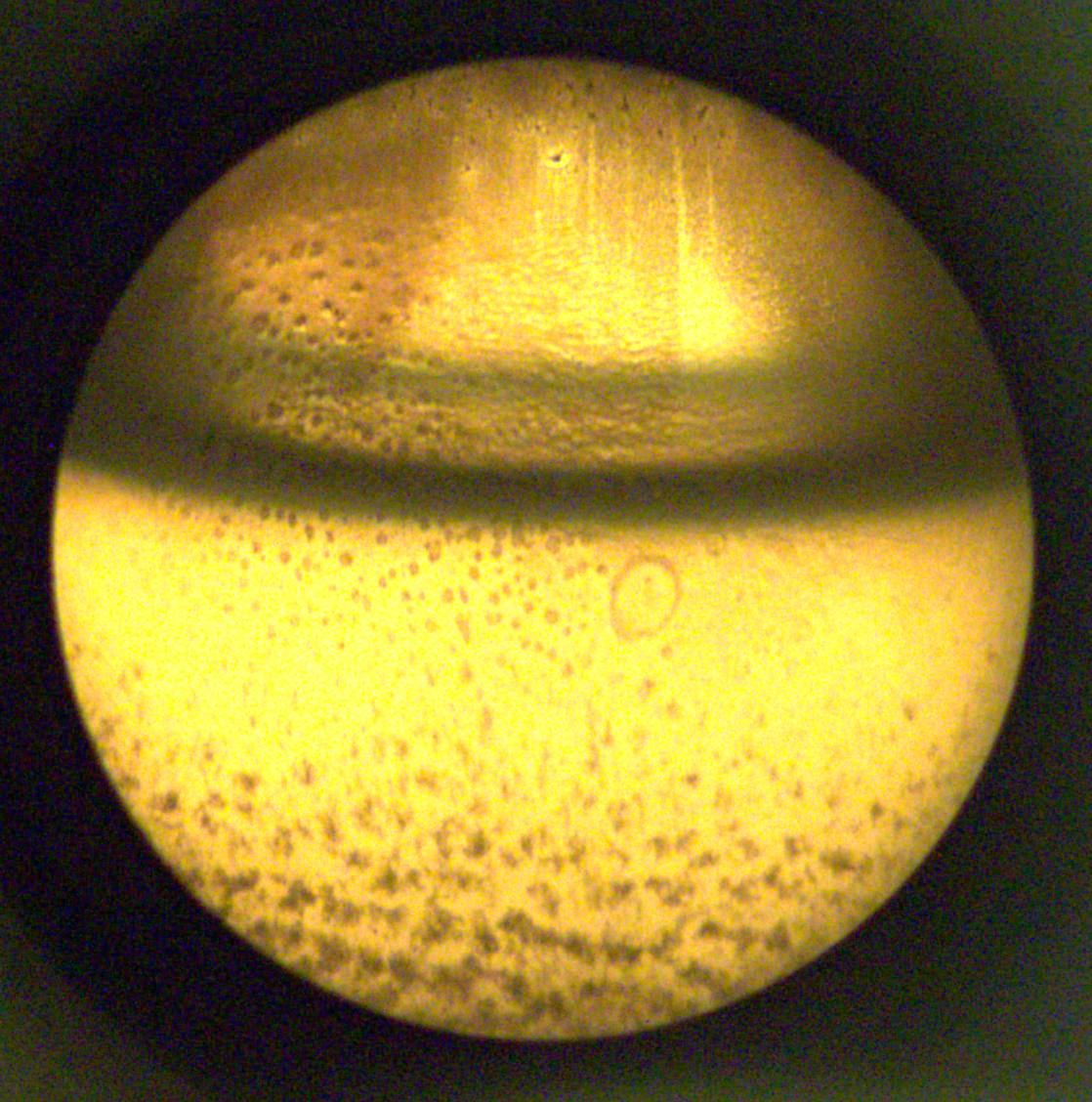}
    \caption{Liquid \HTwo{} with nitrogen impurities.}
  \label{fig:cell_photographs:with_nitrogen_impurities}
\end{subfigure}%
\hspace{1em}
\begin{subfigure}{.3\textwidth}
  \centering
    \includegraphics[width=\linewidth]{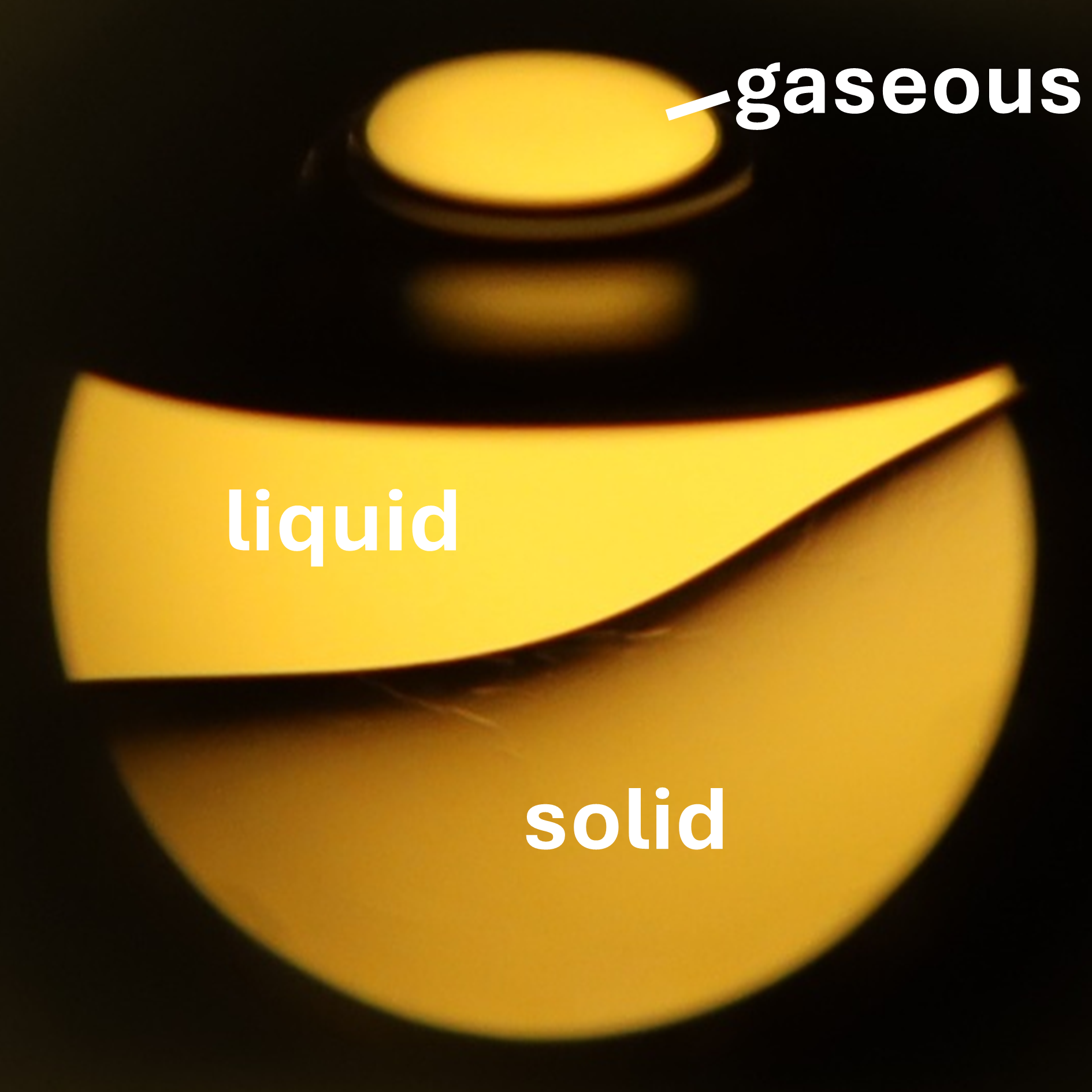}
    \caption{Coexistence of the three phases of \HTwo{}.}
  \label{fig:cell_photographs:three_phases}
\end{subfigure}%
\hspace{1em}
\begin{subfigure}{.3\textwidth}
  \centering
    \includegraphics[width=\linewidth]{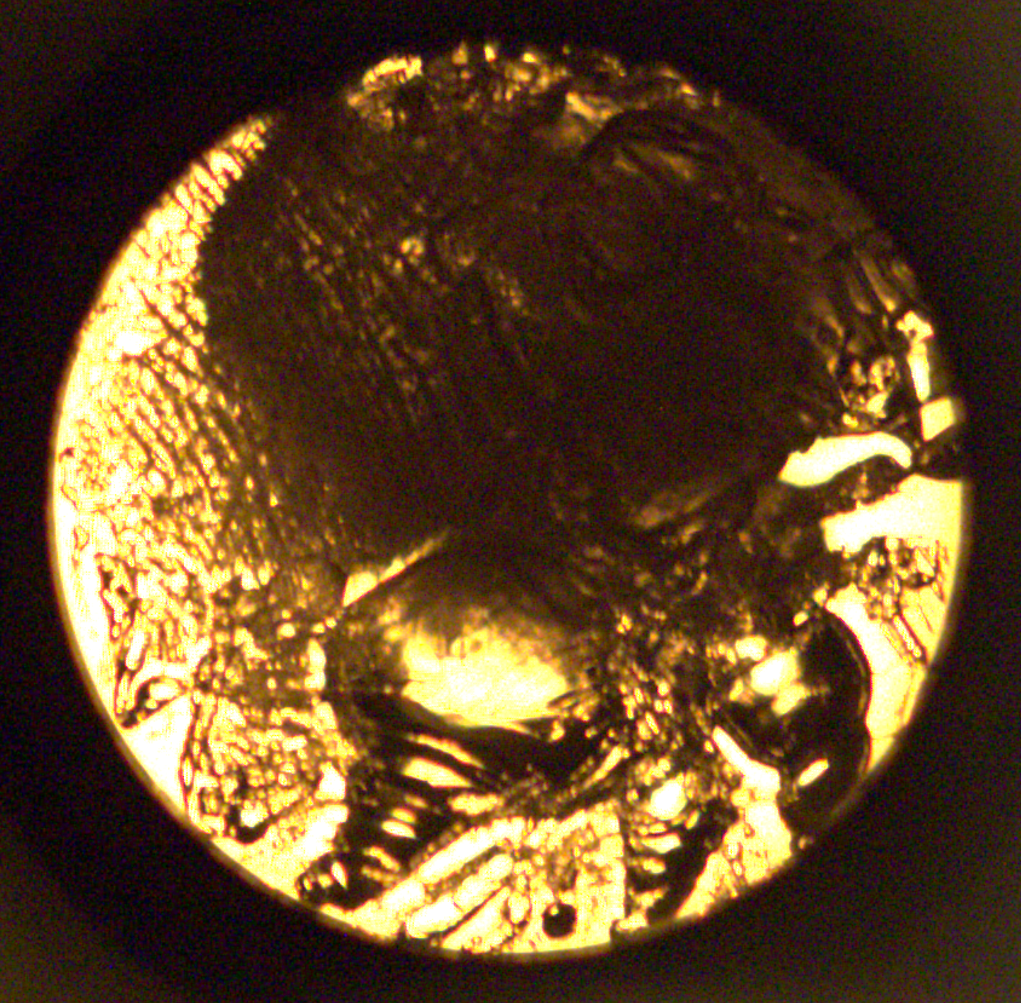}
    \caption{Cracked solid of quickly frozen \HTwo{}.}
  \label{fig:cell_photographs:cracked_crystal}
\end{subfigure}%
\caption{Photographs of \HTwo{} in the measurement cell taken with warm white back-illumination.}
  \label{fig:cell_photographs}
\end{figure}

Overall, cooldown and warmup of the cryostat proved to be non-critical.
The only point of note being a pressure rise in the insulation vacuum upon warmup, which is, however, not an issue for safe operation. 
One important insight early on was the importance of gas purity. 
Tiny admixtures of impurities, in particular nitrogen, can form tiny, snow-like particles when the gas is filled into the measurement cell at the temperature range of the liquid hydrogen isotopologues (\SIrange{\approx10}{30}{\kelvin}). 
This nitrogen snow settles in the measurement cell, including on the windows as shown in \cref{fig:cell_photographs:with_nitrogen_impurities}.
As it is opaque and appears to be brownish in color, it can potentially hinder the spectroscopic investigations.
To remove the nitrogen snow, the liquefied \HTwo{}, \DTwo{}, or Ne is evaporated by increasing the temperature slightly above their boiling point (\SI{\approx 35}{\kelvin}) and pumping of the measurement cell contents into a buffer vessel, leaving behind the impurities. 
This is followed by warming up the measurement cell to around \SI{80}{\kelvin}, at which point the snow appears to quickly sublimate and the resulting gas can be pumped off. 
The pure gas previously stored in a buffer vessel can then be liquefied again without having a relevant amount of impurities present. 

When pure \HTwo{} gas is liquefied and slowly brought to a temperature around its triple point, the liquid phase begins to solidify, leading to a 3-phase system inside the measurement cell as can be seen in \cref{fig:cell_photographs:three_phases}. 
Solid \HTwo{} forms the bottom layer, atop which a layer of liquid \HTwo{} and then gaseous \HTwo{} is present. 
The surface tension between the liquid and gaseous \HTwo{} can be seen from the dark meniscus, where the light of the back-illumination is refracted away from the camera.
A similar dark edge is visible between the solid and liquid layers due to the boundary not being perpendicular to the viewing axis. 
On this dark edge, tiny stripes of lighter color are visible, which are the result of tiny cracks in the solid refracting the light differently.
These structures can be made much visible by including linear polarization filters, rotated against each other by \SI{90}{\degree}, in front and behind of the measurement cell.
The polarization filters remove the light shining through the gas and liquid phase, but light passing through the solid \HTwo{} has its plane of polarization affected depending on the thickness of the solid and stresses therein, making these areas much brighter and giving more contrast to changes in the solid structure. 

When pure \HTwo{} gas is liquefied and solidified by cooling it down quickly below its freezing point, the solid goes through a stage where fronts of melting and re-solidification repeatedly move across the measurement cell, creating many internal surfaces, which cause light to be refracted to the point that the entire measurement cell appears opaque. 
An example of this is shown in \cref{fig:cell_photographs:cracked_crystal}.
In this state, no transmission spectroscopy can be performed. 
This opaque solidification is only temporary, as, even when keeping the \HTwo{} at temperatures below its triple point, an annealing of the \HTwo{} solid is observed, turning the opaque solid homogeneous and transparent on the timescale of several hours to one day.

\section{Conclusion}
\label{sec:conclusion}

The design, manufacturing and inactive commissioning of the \gls{tapir2} experiment has been completed.
We have managed to fulfill the challenging combined requirements including high amounts of tritium (\requiredTritiumAmountGram{}), overpressure of up to \operatingPressureLimit{}, covering all three phases of all six hydrogen isotopologues down to a temperature of \achievedTemperatureLowerLimitKelvin{}, and doing so in a cell accessible to optical and spectral analytic tools in a tightly constrained space.
All of this is accomplished while maintaining the \gls{tlk} safety philosophy, ensuring safe tritium operation.
With full integration into the \gls{tlk} closed loop tritium infrastructure at hand, the \gls{tapir2} experiment is ready for tritium operation. 
\section{Outlook}
\label{sec:outlook}

Currently, a thorough inactive (without tritium \TTwo{}) scientific commissioning campaign with protium (\HTwo{}) and deuterium (\DTwo{}) is running for further functionality tests, optimization of the optical setup, and first experimental results. 
In these measurements, all experiments which would be affected by residual tritium inside the apparatus, and can therefore not be conducted after commissioning with tritium, will be performed.
This covers for example the natural \orthopara{} conversion in the setup as a reference for future measurements of radiolytically driven conversion induced by the \betadecay{} of tritium.

After a successful finalization of the inactive campaign, the active commissioning with first tritium operations will be started in 2026. 
After successful commissioning of the secondary containment enclosure as well as measurement devices relying on effects of the tritium \betadecay{}, such as ionization chambers, scientific measurements will commence. 
These measurements will then focus on several key goals: radiolytic \orthopara{} conversion close to the triple point and at high pressure to study the thermodynamic behavior of \QTwo{}, phase change behavior of tritium and tritiated mixtures, and calibration of the optical measurement methods for these mixtures. 
The latter in particular is one of the last big missing steps of an analytic system for isotope separation systems in the fusion fuel cycle.

\section*{Acknowledgments} 
 
 This work was partially supported by the Bundesministerium für For-schung, Technologie und Raumfahrt (BMFTR, German Federal Ministry of Research, Technology and Space) within the \textit{Verbundprojekt: Inertial Fusion Energy (IFE) Targetry HUB für die DT-Trägheitsfusion (IFE-Targetry-HUB)} - \textit{Teilvorhaben: Tritium-Phasenraum-Navigation (TriPaN)} with grant number 13F1013H.
 The authors acknowledge the support of the TLK members, especially Nancy Tuchscherer, Tobias Falke, Tobias Weber, and Adalbert Braun.

\printnoidxglossaries

\bibliographystyle{elsarticle-num} 
\bibliography{bibliography}

\end{document}